# Perfectly Normal Type-2 Fuzzy Interpolation B-spline Curve

**Rozaimi Zakaria**

Department of Mathematics, Faculty of Science and Technology, Universiti Malaysia Terengganu, Malaysia.
rozaimi_z@yahoo.com

**Abd. Fatah Wahab**

Department of Mathematics, Faculty of Science and Technology, Universiti Malaysia Terengganu, Malaysia.
fatah@umt.edu.my

**R. U. Gobithaasan**

Department of Mathematics, Faculty of Science and Technology, Universiti Malaysia Terengganu, Malaysia.
gr@umt.edu.my

**Abstract**

In this paper, we proposed another new form of type-2 fuzzy data points(T2FDPs) that is perfectly normal type-2 data points(PNT2FDPs). These kinds of brand-new data were defined by using the existing type-2 fuzzy set theory(T2FST) and type-2 fuzzy number(T2FN) concept since we dealt with the problem of defining complex uncertainty data. Along with this restructuring, we included the fuzzification(alpha-cut operation), type-reduction and defuzzification processes against PNT2FDPs. In addition, we used interpolation B-soline curve function to demonstrate the PNT2FDPs.

## 1 Introduction

In dealing the problem of defining the uncertainty, there are various methods can be applied for dealing those problems. A well-known method in define the uncertainty problem is type-1 fuzzy set theory(fuzzy set theory) has recently involved in many areas of interest which, introduced by Zadeh in 1965 [31]. By the type-1 fuzzy set theory(T2FST) basic definition, then there also appear many basic concepts, which dealing the problem of uncertainty in specific areas such as type-1 fuzzy number(T1FN) concept, which deals to define the uncertainty real data problem.

Although T1FST was being the method of dealing the uncertainty problem, there also exist a problem when dealing the complex uncertainty. Therefore, Zadeh introduced the next level of T1FST, that is T2FST [32]. This theory used to solve the complex uncertainty problem, which is also can be used to solve the uncertainty problem.

The set of collection's data points are very important to perform curve and surface as the representative of the set data points' nature. When these data points become uncertain, then both curve and surface are unable to represent for those uncertain data points. Therefore, by using the T1FST and T1FN, then the uncertainty data points can be defined, which known as type-1 fuzzy data points(T1FDPs) and also are able to construct curve and surface [1-8,14,21,23,24,26,27,25,28,29].

However, in dealing the problem in defining the complex uncertainty data, we used the T2FST, T2FN concept and type-2 fuzzy relation(T2FR) to solve this matter. After the complex uncertainty data had been defined and become T2FDPs, then we applied fuzzification(alpha-cut operation), type-reduction and defuzzification process in order getting the crisp type-2 fuzzy solution data points(CT2FSDPs) as the final result.

Therefore, in this paper, we will discuss about what the PNT2FDPs was and how to define the PNT2FDPs then. By defining the PNT2FDPs, then we applied the fuzzification, type-reduction and defuzzification processes, which will be modeled by interpolation B-spline curve function.

This paper organized as follows: Section 2 discusses of some basics definitions of T2FST. Then, in Section 3, we discuss about the T2FDPs definition along with the alpha-cut operation process, type-reduction and also defuzzification methods. For Section 4, we will discuss about the development of PNT2FDPs based on the definition of PNT2TFN and the processes of getting CT2FSDPs before we demonstrate for all these processes that had been mentioned before by using the interpolation B-spline curve function as the hypothetical example in Section 5. Then, this proposed method can be concluded in Section 6.

## 2 Type-2 Fuzzy Set Theory

In this section, we will define some basic theories of T2FST, which given as follows.

**Definition 2.1.** A type-2 fuzzy set(T2FS), denoted $\ddot{\vec{A}}$, is characterized by a

type-2 membership function $\mu_{\overset{\leftrightarrow}{A}}(x,u)$, where $x \in X$ and $u \in U_x \subseteq [0,1]$ that is,
$\overset{\leftrightarrow}{A} = \left\{ \left( (x,u), \mu_{\overset{\leftrightarrow}{A}}(x,u) \right) \mid \forall x \in X, \forall u \in U_x \subseteq [0,1] \right\}$
in which, $0 \le \mu_{\overset{\leftrightarrow}{A}}(x,u) \le 1$ [22].

**Definition 2.2.** A T2FN is broadly defined as a T2FS that has a numerical domain. An interval T2FS is defined using the following four constraints, where $\overset{\leftrightarrow}{A}_\alpha = \left\{ [a^\alpha, b^\alpha], [c^\alpha, d^\alpha] \right\}$, $\forall \alpha \in [0,1]$, $\forall a^\alpha, b^\alpha, c^\alpha, d^\alpha \in \mathbb{R}$ (Fig. 2.1) [19,33]:

1. $a^\alpha \le b^\alpha \le c^\alpha \le d^\alpha$
2. $[a^\alpha, d^\alpha]$ and $[b^\alpha, c^\alpha]$ generate a function that is convex and $[a^\alpha, d^\alpha]$ generate a function is normal.
3. $\forall \alpha_1, \alpha_2 \in [0,1] : (\alpha_2 > \alpha_1) \Rightarrow \left( \left[ a^{\alpha_1}, c^{\alpha_1} \right] \supset \left[ a^{\alpha_2}, c^{\alpha_2} \right], \right.$
   $\left. \left[ b^{\alpha_1}, d^{\alpha_1} \right] \supset \left[ b^{\alpha_2}, d^{\alpha_2} \right] \right)$, for $c^{\alpha_2} \ge b^{\alpha_2}$.
4. If the maximum of the membership function generated by $[b^\alpha, c^\alpha]$ is the level $\alpha_m$, that is, $[b^{\alpha_m}, c^{\alpha_m}]$, then $\left[ b^{\alpha_m}, c^{\alpha_m} \right] \subset \left[ a^{\alpha=1}, d^{\alpha=1} \right]$.

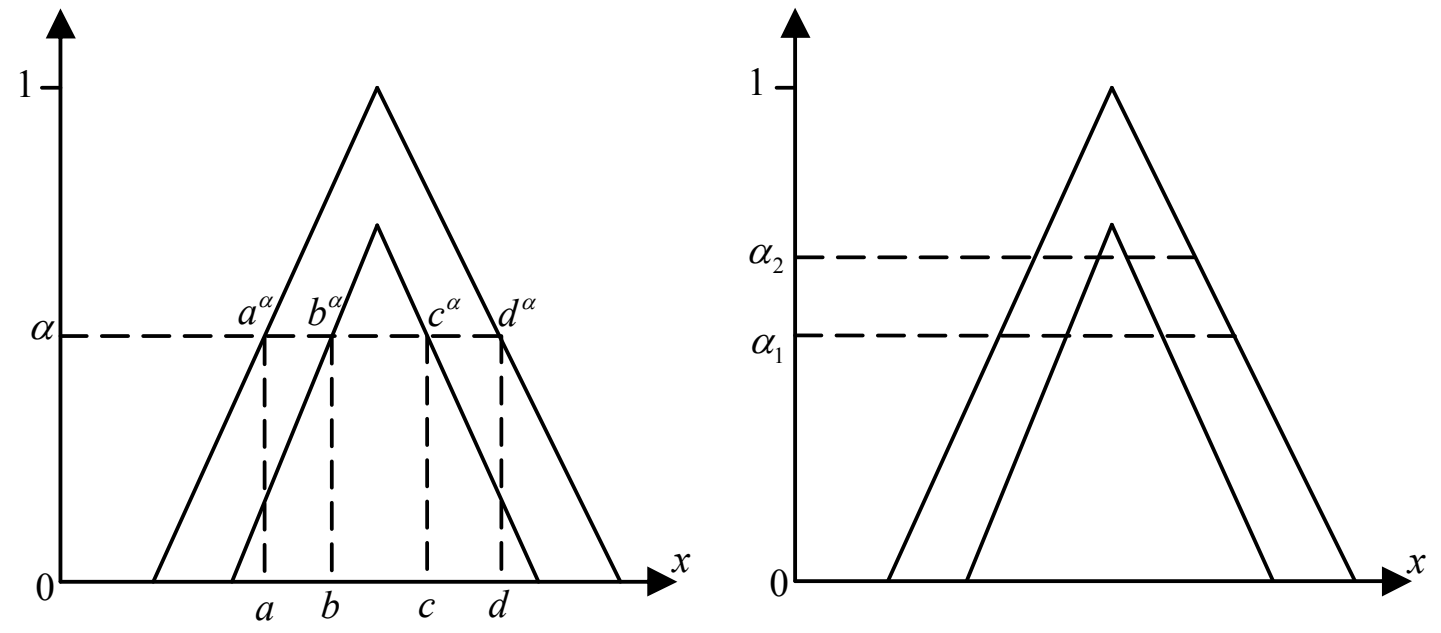

Figure 2.1. Definition of an interval T2FN.

**Definition 2.3.** Let $X, Y, U_x, V_y \subseteq R$ and

$\overset{\leftrightarrow}{A} = \left\{ \left( (x,u), \mu_{\overset{\leftrightarrow}{A}}(x,u) \right) \mid \forall x \in X, \forall u \in U_x \subseteq [0,1] \right\}$ and

$\overset{\leftrightarrow}{B} = \left\{ \left( (y,v), \mu_{\overset{\leftrightarrow}{B}}(y,v) \right) \mid \forall y \in Y, \forall v \in V_y \subseteq [0,1] \right\}$

are two T2FSs. Then, $\overset{\leftrightarrow}{R} = \left\{ \left( \left( (x,u), (y,v) \right), \right. \right.$ $\left. \left. \mu_{\overset{\leftrightarrow}{R}} \left( \mu_{\overset{\leftrightarrow}{A}}(x,u), \mu_{\overset{\leftrightarrow}{B}}(y,v) \right) \right) \mid \left( \forall x \in X, \forall u \in U_x \right) \times \left( \forall y \in Y, \forall u \in V_y \right) \subseteq [0,1] \right\}$ is a T2FR on $\overset{\leftrightarrow}{A}$ and $\overset{\leftrightarrow}{B}$ if $\mu_{\overset{\leftrightarrow}{R}} \left( \mu_{\overset{\leftrightarrow}{A}}(x,u), \mu_{\overset{\leftrightarrow}{B}}(y,v) \right) \le \mu_{\overset{\leftrightarrow}{A}}(x,u)$ , $\forall \left( (x,u), (y,v) \right) \in \left( \forall x \in X, \forall u \in U_x \right) \times$

$\left(\forall y \in Y, \forall v \in V_y\right)$.

## 3 Type-2 Fuzzy Data Points

For this section, we define the T2FDPs based on the definitions of T2FST, T2FN and T2FR early had been defined.

**Definition 3.1.** Let $P = \{x \mid x \text{ type-2 fuzzy point}\}$ and $\overset{\leftrightarrows}{P} = \{P_i \mid P_i \text{ data point}\}$ which is set of type-2 fuzzy data point with $P_i \in P \subset X$, where $X$ is a universal set and $\mu_P(P_i): P \to [0,1]$ is the membership function which defined as $\mu_P(P_i) = 1$ and formulated by $\overset{\leftrightarrows}{P} = \{(P_i, \mu_P(P_i)) \mid P_i \in \square\}$ . Therefore,

$$\mu_P(P_i) = \begin{cases} 0 & \text{if } P_i \notin X \\ c \in (0,1) & \text{if } P_i \ddot{\in} X \\ 1 & \text{if } P_i \in X \end{cases} \tag{3.1}$$

with $\mu_P(P_i) = \left\langle \mu_P(\overset{\leftrightarrows}{P}_i^{\leftarrow}), \mu_P(P_i), \mu_P(\overset{\leftrightarrows}{P}_i^{\rightarrow}) \right\rangle$ which $\mu_P(\overset{\leftrightarrows}{P}_i^{\leftarrow})$ and $\mu_P(\overset{\leftrightarrows}{P}_i^{\rightarrow})$ are left and right footprint of membership values with $\mu_P(\overset{\leftrightarrows}{P}_i^{\leftarrow}) = \left\langle \mu_P(\overset{\leftarrow}{\vec{P}_i^{\leftarrow}}), \mu_P(\vec{P}_i^{\leftarrow}), \mu_P(\overset{\rightarrow}{\vec{P}_i^{\leftarrow}}) \right\rangle$ where, $\mu_P(\overset{\leftarrow}{\vec{P}_i^{\leftarrow}})$, $\mu_P(\vec{P}_i^{\leftarrow})$ and $\mu_P(\overset{\rightarrow}{\vec{P}_i^{\leftarrow}})$ are left-left, left, right-left membership grade values and $\mu_P(\overset{\rightarrow}{\vec{P}_i^{\rightarrow}})$, $\mu_P(\vec{P}_i^{\rightarrow})$ and $\mu_P(\overset{\leftarrow}{\vec{P}_i^{\rightarrow}})$ are right-right, right, left-right membership grade values, which can be written as

$$\overset{\leftrightarrows}{P} = \left\{\overset{\leftrightarrows}{P}_i : i = 0,1,2,\dots,n\right\} \tag{3.2}$$

for every $i$, $\overset{\leftrightarrows}{P}_i = \left\langle \overset{\leftrightarrows}{P}_i^{\leftarrow}, P_i, \overset{\leftrightarrows}{P}_i^{\rightarrow} \right\rangle$ with $\overset{\leftrightarrows}{P}_i^{\leftarrow} = \left\langle \overset{\leftarrow}{\vec{P}_i^{\leftarrow}}, \vec{P}_i^{\leftarrow}, \overset{\rightarrow}{\vec{P}_i^{\leftarrow}} \right\rangle$ where $\overset{\leftarrow}{\vec{P}_i^{\leftarrow}}$, $\vec{P}_i^{\leftarrow}$ and $\overset{\rightarrow}{\vec{P}_i^{\leftarrow}}$ are left-left, left and right-left T2FDPs and $\overset{\leftrightarrows}{P}_i^{\rightarrow} = \left\langle \overset{\leftarrow}{\vec{P}_i^{\rightarrow}}, \vec{P}_i^{\rightarrow}, \overset{\rightarrow}{\vec{P}_i^{\rightarrow}} \right\rangle$ where $\overset{\leftarrow}{\vec{P}_i^{\rightarrow}}$, $\vec{P}_i^{\rightarrow}$ and $\overset{\rightarrow}{\vec{P}_i^{\rightarrow}}$ are left-right, right and right-right T2FDPs respectively. This can be illustrated as in Fig. 3.1.

The illustration of T2FDP was shown in Fig. 3.1 which T1FDP becomes the primary membership function bounded by upper bound, $\left[\overset{\leftarrow}{\vec{P}^{\leftarrow}}, P, \overset{\rightarrow}{\vec{P}^{\rightarrow}}\right]$ and lower bound, $\left[\overset{\rightarrow}{\vec{P}^{\leftarrow}}, P, \overset{\leftarrow}{\vec{P}^{\rightarrow}}\right]$ respectively. The process of defining T2FDP can be shown through Fig. 3.2.

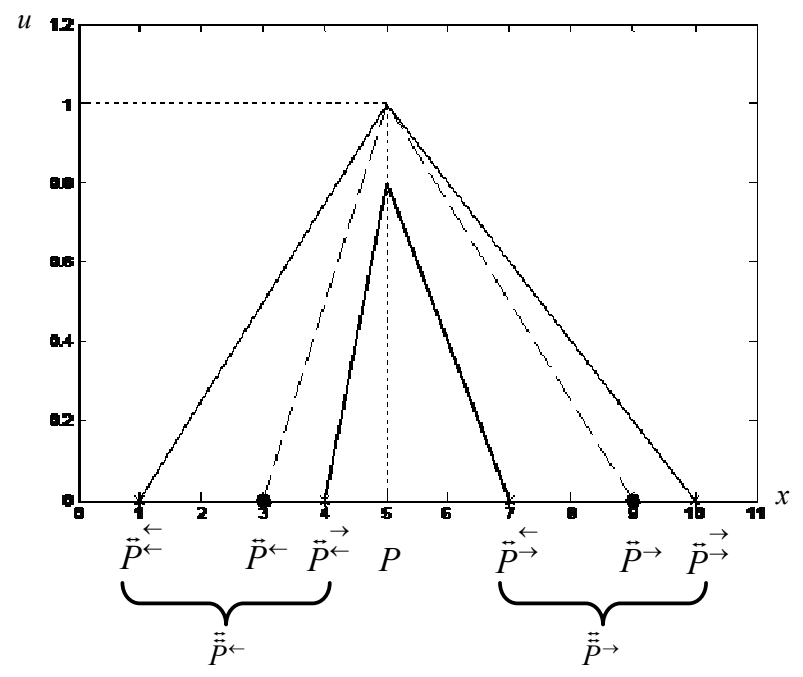

Figure 3.1. T2FDP around 5.

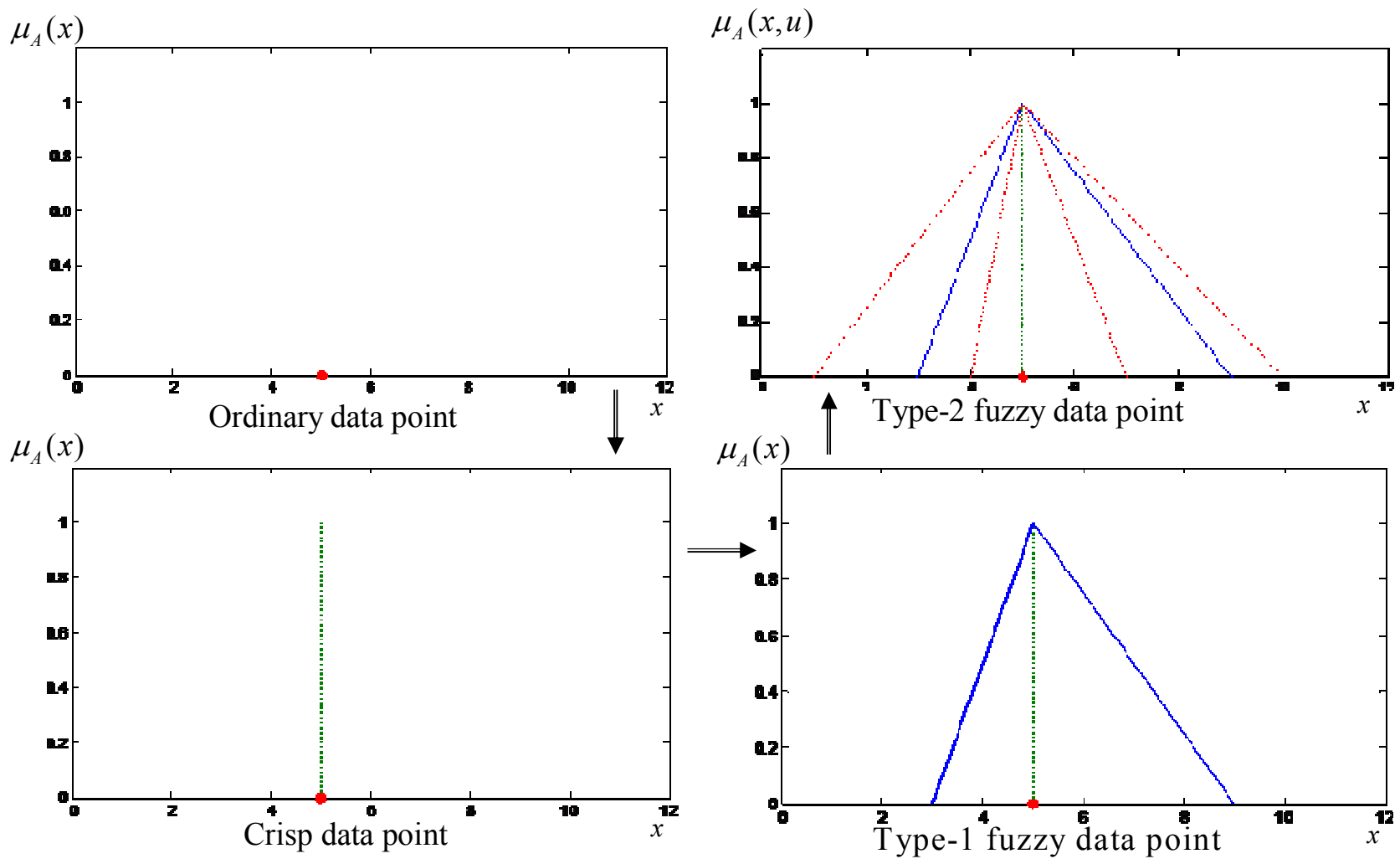

Figure 3.2. Process of defining T2FDP.

Fig. 3.2 shows that the process of defining T2FDP from the ordinary point. This T2FDP formed based on the definition of T2FN and T2FR.

## 4 Perfectly Normal Type-2 Fuzzy Data Points

In this Section 4, we will discuss about the definition of PNT2FDP along with its fuzzification process(alpha-cut operation), type-reduction and defuzzification methods.

**Definition 4.1.** Given that T2FDP, $\ddot{\vec{P}}$ which the height of LMF and UMF are $h\left(\underset{\smile}{\ddot{\vec{P}}}\right)$ and $h\left(\underset{\frown}{\ddot{\vec{P}}}\right)$ respectively, then T2FDP called **Perfectly Normal** of T2FDP(PNT2FDP) if $h\left(\underset{\smile}{\ddot{\vec{P}}}\right)=h\left(\underset{\frown}{\ddot{\vec{P}}}\right)=1$ [18]. This Def. 4.1 can be illustrated through Fig. 4.1.

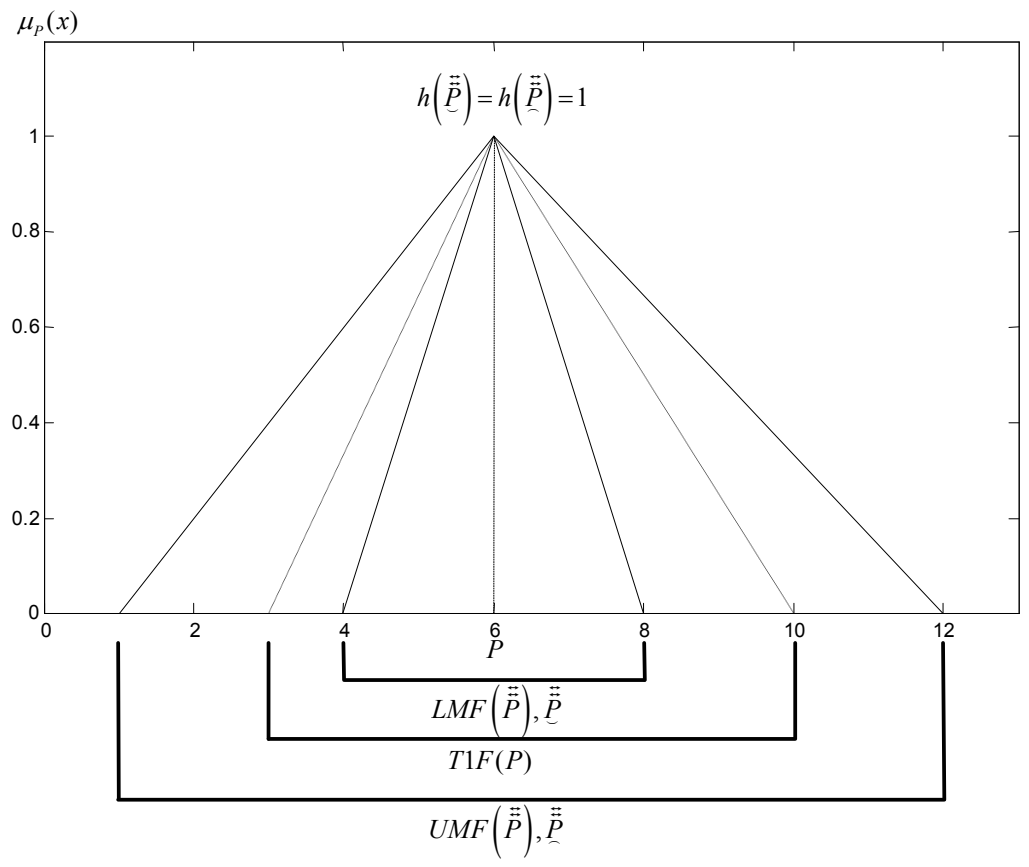

Figure 4.1. The PNT2FDPs.

**Definition 4.2.** Based on Def. 3.1, let $\ddot{\vec{P}}$ be the set of T2FDPs with $\ddot{\vec{P}}_i \in \ddot{\vec{P}}$ where $i=0,1,...,n-1$. Then ${}^{PN}\ddot{\vec{P}}^{\alpha}$ is the $\alpha$-cut operation of PNT2FDPs which is given as equation as follows.

$$
\begin{aligned}
{}^{PN}\ddot{\vec{P}}_i^{\alpha} &= \left\langle {}^{PN}\ddot{\vec{P}}_i^{\alpha\leftarrow}, P_i, {}^{PN}\ddot{\vec{P}}_i^{\alpha\rightarrow} \right\rangle \\
&= \left\langle \left\langle \overset{\leftarrow}{\vec{P}}{}_i^{\alpha\leftarrow}; \vec{P}_i^{\alpha\leftarrow}; \overset{\rightarrow}{\vec{P}}{}_i^{\alpha\leftarrow} \right\rangle, P_i, \left\langle \overset{\leftarrow}{\vec{P}}{}_i^{\alpha\rightarrow}; \vec{P}_i^{\alpha\rightarrow}; \overset{\rightarrow}{\vec{P}}{}_i^{\alpha\rightarrow} \right\rangle \right\rangle \\
&= \left\langle \left[ \left( P_i - \left\langle \overset{\leftarrow}{\vec{P}}{}_i^{\leftarrow}; \vec{P}_i^{\leftarrow}; \overset{\rightarrow}{\vec{P}}{}_i^{\leftarrow} \right\rangle \right)\alpha + \left\langle \overset{\leftarrow}{\vec{P}}{}_i^{\leftarrow}; \vec{P}_i^{\leftarrow}; \overset{\rightarrow}{\vec{P}}{}_i^{\leftarrow} \right\rangle \right], P_i, \right. \\
&\quad \left. \left[ -\left( \left\langle \overset{\leftarrow}{\vec{P}}{}_i^{\rightarrow}; \vec{P}_i^{\rightarrow}; \overset{\rightarrow}{\vec{P}}{}_i^{\rightarrow} \right\rangle - P_i \right)\alpha + \left\langle \overset{\leftarrow}{\vec{P}}{}_i^{\rightarrow}; \vec{P}_i^{\rightarrow}; \overset{\rightarrow}{\vec{P}}{}_i^{\rightarrow} \right\rangle \right] \right\rangle
\end{aligned}
\tag{4.1}
$$

This definition can be illustrated through Fig. 4.2.

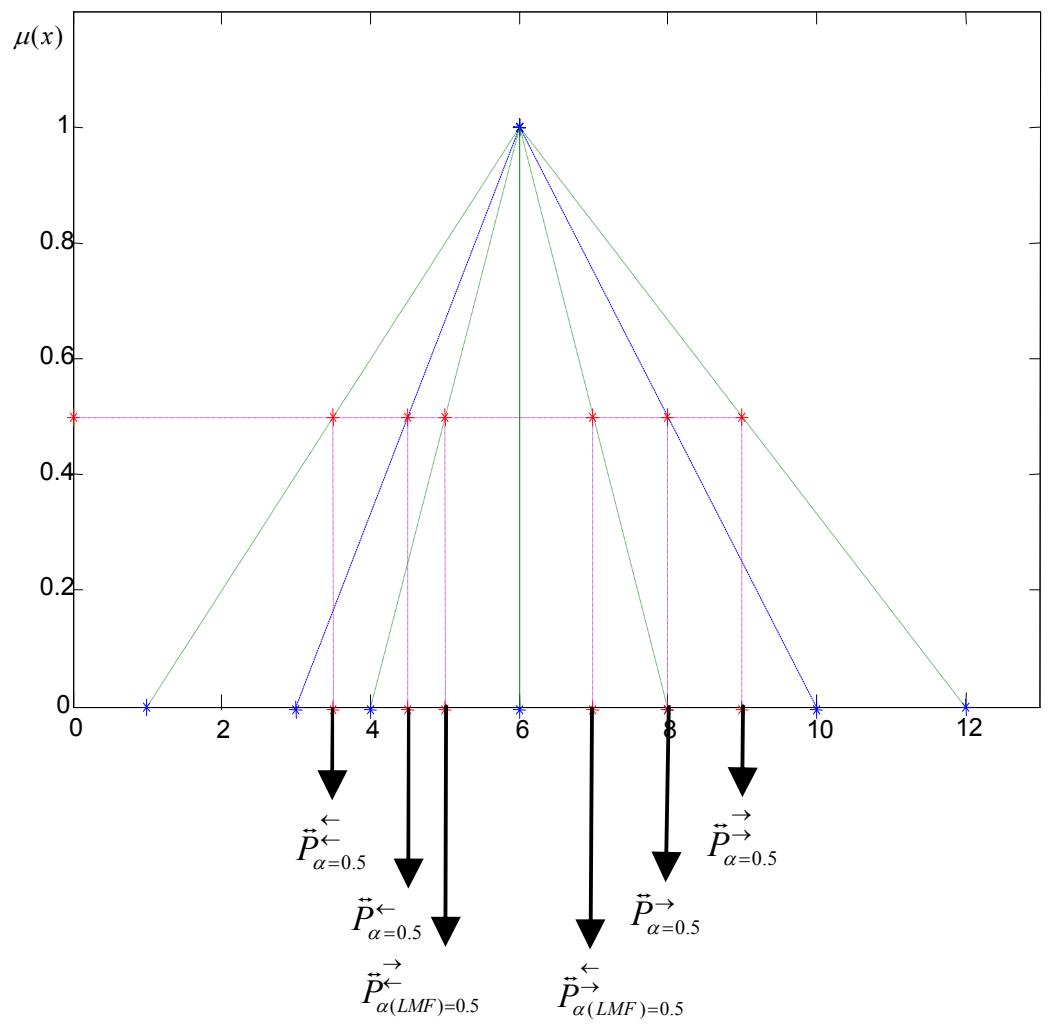

Figure 4.2. The alpha-cut operation towards PNT2FDP.

After the fuzzification process against PNT2FDPs had been applied, then the next step is the type-reduction method. Type-reduction method is a method were used to reduce T2FDPs to T1FDPs after fuzzificication process has been applied. In addition, type-reduction used to allow defuzzification method of type-1 can be applied. There are many types of type-reduction, which has mentioned in [10,11,13,15-17]. However, in this paper, we construct another type-reduction method, which is known as centroid min method, which can be given via Def. 4.3.

**Definition 4.3.** Let ${}^{PN}\overleftrightarrow{P}_i$ be a set $(n+1)$ PNT2FDPs, then type-reduction method of $\alpha$ -T2FDPs(after fuzzification), ${}^{PN}\overline{\overleftrightarrow{P}}_i$ is defined by

$$ {}^{PN}\overline{\overleftrightarrow{P}}^{\alpha} = \left\{ {}^{PN}\overline{\overleftrightarrow{P}}_i^{\alpha} = \left\langle {}^{PN}\overline{\overleftrightarrow{P}}_i^{\alpha\leftarrow}, P_i, {}^{PN}\overline{\overleftrightarrow{P}}_i^{\alpha\rightarrow} \right\rangle ; \, i = 0,1,\ldots,n \right\} \tag{4.3} $$

where ${}^{PN}\overline{\overleftrightarrow{P}}_i^{\alpha\leftarrow}$ is left type-reduction of $\alpha$ -cut PNT2FDPs, ${}^{PN}\overline{\overleftrightarrow{P}}_i^{\alpha\leftarrow} = \frac{1}{3}\sum_{i=0,\ldots,n} \left\langle \overleftarrow{\vec{P}}_i^{\alpha\leftarrow} + \vec{P}_i^{\alpha\leftarrow} + \overrightarrow{\vec{P}}_i^{\alpha\leftarrow} \right\rangle$, $P_i$ is the crisp point and ${}^{PN}\overline{\overleftrightarrow{P}}_i^{\alpha\rightarrow}$ is right type-reduction of $\alpha$ -cut PNT2FDPs, ${}^{PN}\overline{\overleftrightarrow{P}}_i^{\alpha\rightarrow} = \frac{1}{3}\sum_{i=0,\ldots,n} \left\langle \overleftarrow{\vec{P}}_i^{\alpha\rightarrow} + \vec{P}_i^{\alpha\rightarrow} + \overrightarrow{\vec{P}}_i^{\alpha\rightarrow} \right\rangle$.

**Definition 4.4.** Let $\alpha$ -TR is the type-reduction method after $\alpha$ -cut process had been applied for every PNT2FDPs, ${}^{PN}\overline{\overleftrightarrow{P}}_i^{\alpha}$. Then ${}^{PN}\overline{\overline{P}}_i^{\alpha}$ named as defuzzification PNT2FDPs for ${}^{PN}\overline{\overleftrightarrow{P}}_i^{\alpha}$ if for every ${}^{PN}\overline{\overleftrightarrow{P}}_i^{\alpha} \in {}^{PN}\overline{\overleftrightarrow{P}}^{\alpha}$,

$$ {}^{PN}\overline{\overline{P}}^{\alpha} = \left\{ {}^{PN}\overline{\overline{P}}_i^{\alpha} \right\} \quad \text{for} \quad i = 0,1,\ldots,n \tag{4.4} $$

where for every ${}^{PN}\overline{\overline{P}}_i^{\alpha} = \frac{1}{3}\sum_{i=0} < {}^{PN}\overline{\overline{P}}_i^{\alpha\leftarrow}, P_i, {}^{PN}\overline{\overline{P}}_i^{\alpha\rightarrow} >$. The process in defuzzifying PNT2FDPs can be illustrated at Fig. 4.4.

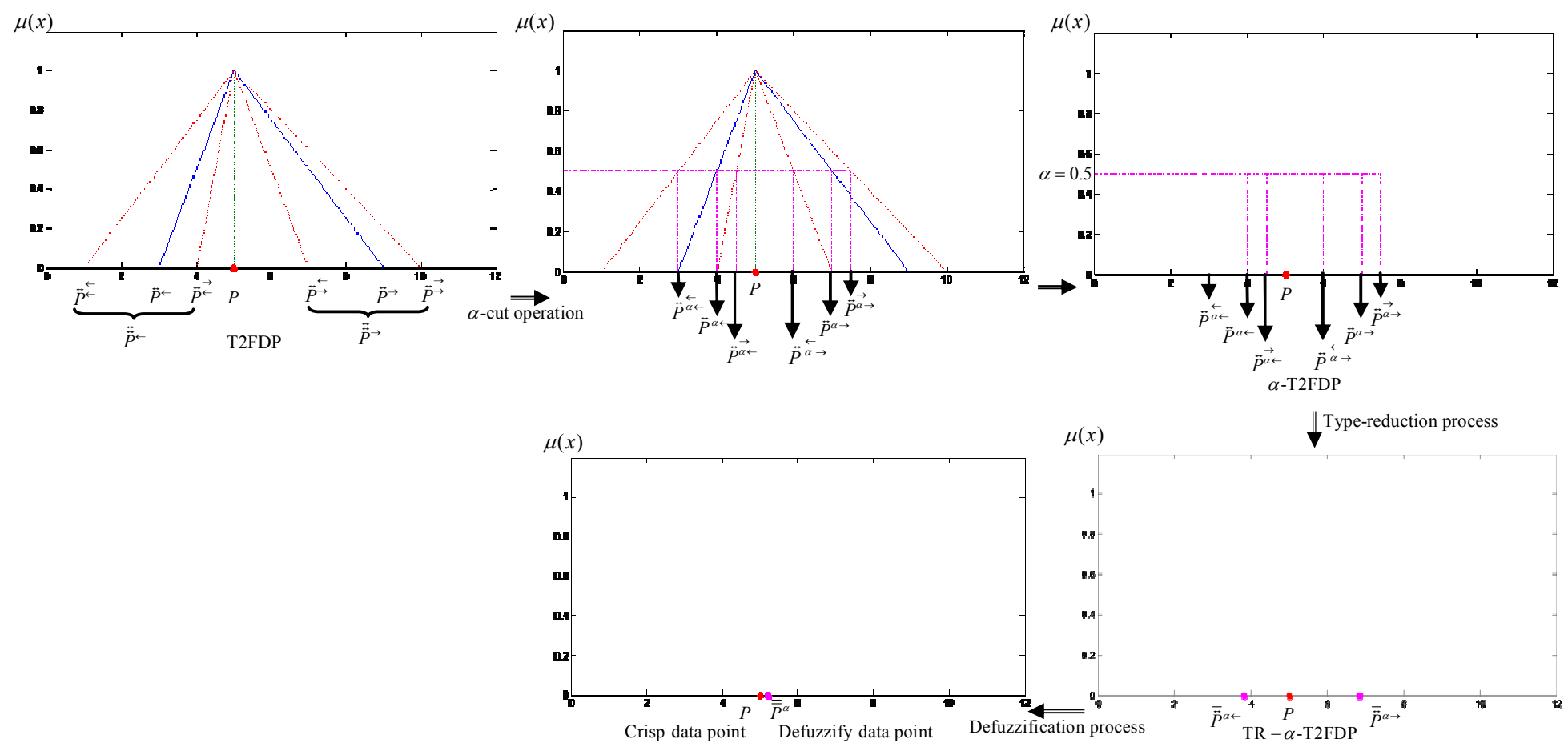

Figure 4.4. Defuzzification process of PNT2FDP.

## 5 Perfectly Normal Type-2 Fuzzy Interpolation B-spline Curve Modeling

When we finished defined the complex uncertainty data points by using T2FST in PNT2FDPs form, then we want to illustrate them through a curve by interpolation B-spline curve function, and it can be known as perfectly normal type-2 fuzzy interpolation B-spline curve(PNT2FIBsC).

PNT2FIBsC will give us more comprehended in how PNT2FDPs can be modeled with give us the curves shape based on the data points. Besides that, we can understand the meaning of the PNT2FDPs based on the yielded curves. Therefore, as the hypothetical example of modeling the complex uncertainty data points, the process of modeling for PNT2FDPs using interpolation B-spline curve function [9,12,20,23,30] can be given in Def. 5.1 as follows.

**Definition 5.1.** Let ${}^{PN}\overset{\leftrightarrow}{\ddot{d}}_i \in R^m$ be a list of PNT2FDPs with $0 \le i \le n$, then the PNT2FIBsC can be defined as

$$\overleftrightarrow{{}^{PN}BsC}(t) = \sum_{i=0}^{k+h-1} {}^{PN}\overset{\leftrightarrow}{\ddot{P}}_i B_{i,k}(t) \quad \text{which} \quad \overleftrightarrow{{}^{PN}BsC}(t_i) = {}^{PN}\overset{\leftrightarrow}{\ddot{d}}_i, \tag{5.1}$$

where $t$ is crisp knot sequences $t_1, t_2, \ldots, t_{m=k+2(n-1)}$, ${}^{PN}\overset{\leftrightarrow}{\ddot{P}}_i$ are perfectly normal type-2 fuzzy control points(PNT2FCPs) and $B_{i,k}(t)$ is basic function of B-spline. In this part, PNT2FCPs are used to force the constructed curve to interpolate the PNT2FDPs. Here, the PNT2FDPs known as ${}^{PN}\overleftrightarrow{d}_i$.

Therefore, to illustrate the PNT2FIBsC, we summarized in the table form(Table 5.1) for the computed PNT2FDPs and also in curve form by looking at Fig. 5.1 as follows.

Table 5.1. Process of fuzzification, type-reduction and defuzzification of PNT2FBC.

| PNT2FDPs $\overleftrightarrow{{}^{PN}BsC(t)} = {}^{PN}\overleftrightarrow{d}_i$ | ${}^{PN}\overleftrightarrow{d}_i$ | | | | | | |
|---|---|---|---|---|---|---|---|
| | $\overleftrightarrow{d}_i^{\overset{\leftarrow}{\leftarrow}}$ | $\overleftrightarrow{d}_i^{\leftarrow}$ | $\overleftrightarrow{d}_i^{\overset{\rightarrow}{\leftarrow}}$ | $d_i$ | $\overleftrightarrow{d}_i^{\overset{\leftarrow}{\rightarrow}}$ | $\overleftrightarrow{d}_i^{\rightarrow}$ | $\overleftrightarrow{d}_i^{\overset{\rightarrow}{\rightarrow}}$ |
| $i=0$ | (-12, 0) | (-11, 0) | (-9, 0) | (-5, 0) | (3, 0) | (6,0) | (9,0) |
| $i=1$ | (15, 28) | (15, 26) | (15, 25) | (15, 20) | (15, 16) | (15,14) | (15,12) |
| $i=2$ | (17, -13) | (15, -15) | (13, -17) | (10, -20) | (8, -22) | (5,-25) | (3,-27) |
| $i=3$ | (30, 10) | (32, 10) | (34, 10) | (40, 10) | (46, 10) | (48,10) | (49,10) |

| $\alpha$ -Cut, $\alpha = 0.5$ $\overleftrightarrow{{}^{PN}BsC^{\alpha}(t)} = {}^{PN}\overleftrightarrow{d}_i^{\alpha}$ | ${}^{PN}\overleftrightarrow{d}_i^{\alpha}$ | | | | | | |
|---|---|---|---|---|---|---|---|
| | $\overleftrightarrow{d}_i^{\alpha\overset{\leftarrow}{\leftarrow}}$ | $\overleftrightarrow{d}_i^{\alpha\leftarrow}$ | $\overleftrightarrow{d}_i^{\alpha\overset{\rightarrow}{\leftarrow}}$ | $d_i$ | $\overleftrightarrow{d}_i^{\alpha\overset{\leftarrow}{\rightarrow}}$ | $\overleftrightarrow{d}_i^{\alpha\rightarrow}$ | $\overleftrightarrow{d}_i^{\alpha\overset{\rightarrow}{\rightarrow}}$ |
| $i=0$ | (-8.5, 0) | (-8, 0) | (-7, 0) | (-5, 0) | (-1, 0) | (0.5, 0) | (2, 0) |
| $i=1$ | (15, 24) | (15, 23) | (15, 22.5) | (15, 20) | (15, 18) | (15, 17) | (15, 16) |
| $i=2$ | (13.5, -16.5) | (12.5, -17.5) | (11.5, -18.5) | (10, -20) | (9, -21) | (7.5, -22.5) | (6.5, -23.5) |
| $i=3$ | (35, 10) | (36, 10) | (37, 10) | (40, 10) | (43, 10) | (44, 10) | (44.5, 10) |

| Type-Reduction $\overline{\overline{{}^{PN}BsC^{\alpha}}}(t) = {}^{PN}\overleftrightarrow{d}_i^{\alpha}$ | ${}^{PN}\overline{\overline{d}}_i^{\alpha}$ | | |
|---|---|---|---|
| | $\overline{\overline{d}}_i^{\alpha\leftarrow}$ | $d_i$ | $\overline{\overline{d}}_i^{\alpha\rightarrow}$ |
| $i=0$ | (-7.8333, 0) | (-5, 0) | (0.5, 0) |
| $i=1$ | (15, 23.1667) | (15, 20) | (15, 7) |
| $i=2$ | (12.5, -17.5) | (10, -20) | (7.6667, -22.3333) |
| $i=3$ | (36, 10) | (40, 10) | (48.8333, 10) |

| Type-2 Defuzzification $\overline{\overline{{}^{PN}BsC^{\alpha}}}(t) = {}^{PN}\overline{\overline{d}}_i^{\alpha}$ | ${}^{PN}\overline{\overline{d}}_i^{\alpha}$ | |
|---|---|---|
| | $d_i$ | ${}^{PN}\overline{\overline{d}}_i^{\alpha}$ |
| $i=0$ | (-5, 0) | (-4.1111, 0) |
| $i=1$ | (15, 20) | (15, 20.0556) |
| $i=2$ | (10, -20) | (10.0556, -19.9444) |
| $i=3$ | (40, 10) | (39.9444, 10) |

Table 5.1 shows that the numerical example of fuzzification, type-reduction and defuzzification processes of ${}^{PN}\overleftrightarrow{d}_i$ where $i = 0,\ldots,3$. Based on the table, we illustrate all the processes in curve form by using interpolation B-spline curve function as follows.

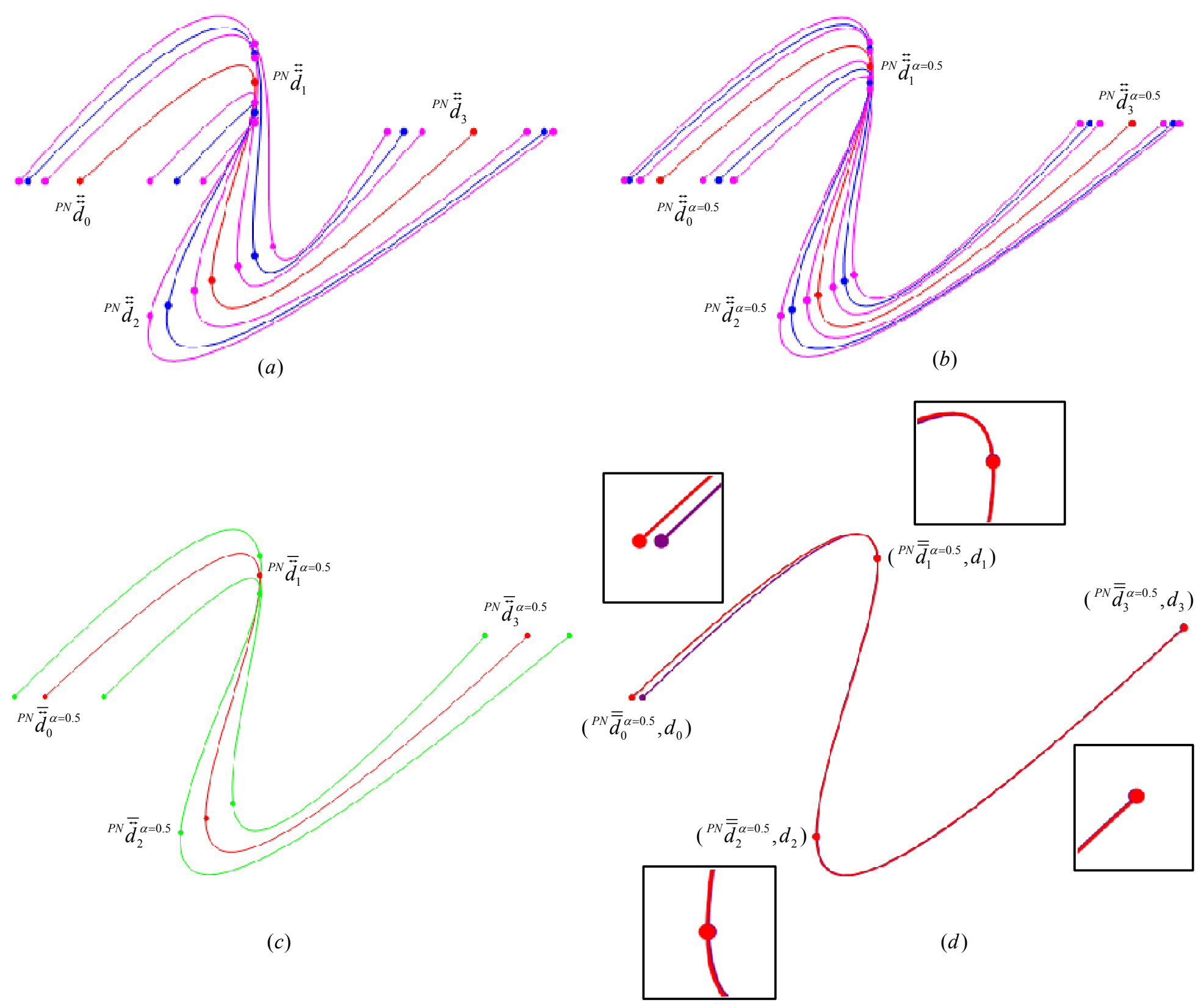

Figure 5.1. The modeling of (a) PNT2FIBsC, (b) $\alpha$ -cut operation of PNT2FIBsC( $\alpha$ -PNT2FIBsC), (c) type-reduction of $\alpha$ -PNT2FIBsC(TR- $\alpha$ -PNT2FIBsC) and (d) defuzzification of TR- $\alpha$ -PNT2FIBsC.

Fig. 5.1 shows that the process of modeling the PNT2FDPs using interpolation B-spline curve function, which become PNT2FIBsC from modeling PNT2FDPs until modeling the defuzzification of PNT2FDPs as a final crisp PNT2FDPs solution modeling.

## 6 Conclusion

In this paper, we constructed PNT2FDP definition based on the concept of T2FN and the definition of T2FR of complex uncertainty data points. The main contribution of this method is to give a new definition of defining the complex uncertainty data which the T1FST unable to define for such data.

When the PNT2FDPs was succeeded defined, we applied some methods to obtain crisp PNT2FDP, which are the fuzzification, type-reduction and defuzzification methods. These methods have their definition to hold the crisp

type-2 fuzzy solution as a final result.

Therefore, in order to illustrate it for more understanding, we blend PNT2FDPs interpolation B-spline Bezier function curve to produce the PNT2FIBsC, which was shown at Table 5.1 as the numerical example and Fig. 5.1 as the modeling curve.

These PNT2FDP can be applied in defining real complex uncertainty data, which can be modeled through various surface functions in approximation and interpolation forms such as Bezier, B-spline, rational Bezier and NURBS functions.

## Acknowledgement

The authors would like to thank Research Management and Innovation Centre (RMIC) of Universiti Malaysia Terengganu and Ministry of Higher Education (MOHE) Malaysia for funding(FRGS, vot59244) and providing the facilities to carry out this research.